\documentclass[aps,prl,twocolumn,showpacs,preprintnumbers,amsmath,amssymb,revsymb]{revtex4-1}
\usepackage{graphicx}
\usepackage{color} 
\usepackage{txfonts}
\usepackage{dcolumn}
\usepackage{bm}
\usepackage{braket}

\definecolor{gray}{rgb}{0.5,0.5,0.5}

\newcommand{\fo}{\mathfrak{f}_{o}}

\def\braket#1{\mathinner{\langle{#1}\rangle}}

\newcommand{\dd}{\textrm{d}} 
\newcommand{\kb}{k_{\textsc{b}}^{}}

\begin{document}

\title{Decoherence of high-energy electrons in weakly
  disordered quantum Hall edge states}
\preprint{}

\author{Simon~E.~Nigg$^{1}$}\email[Corresponding author: ]{simon.nigg@unibas.ch}\author{Anders~Mathias~Lunde$^{2}$}

\affiliation{$^{1}$Department of Physics, University of Basel,
  Klingelbergstrasse 82, 4056 Basel, Switzerland}
\affiliation{$^{2}$Center for Quantum Devices, Niels Bohr Institute, University of Copenhagen, Universitetsparken 5, 2100 Copenhagen, Denmark}
\date{\today}

\begin{abstract}
We investigate theoretically the phase coherence of electron transport in edge states of the integer
quantum Hall effect at filling factor $\nu=2$, in the presence of disorder and inter-edge state Coulomb interaction. Within a Fokker-Planck approach, we calculate analytically the
visibility of the Aharonov-Bohm oscillations of the current through an electronic Mach-Zehnder
interferometer. In agreement with recent
experiments, we find that
the visibility is independent of the energy of the current-carrying electrons
injected high above the Fermi sea. Instead, it is the amount of disorder at the edge that
sets the phase space available for inter-edge state energy exchange
and thereby controls the visibility suppression.
\end{abstract}

\maketitle

Phase coherent electron transport at the edge of a two dimensional
electron gas (2DEG) is a
fascinating topic in condensed matter physics, both because of its
fundamental role in unveiling new correlated states of matter~\cite{Tsui-1982,Wen-1990},
as well as for its practical implications for electronic quantum
information processing~\cite{Sarma-2005, Huynh-2012,Bocquillon-2014,Haack-2014}, and the emerging field of quantum coherent
thermo-electrics~\cite{Hofer-2015,Sanchez-2015}. Although among the
oldest quasi-one dimensional systems to have been
discovered~\cite{Halperin-PRB-1982, Buttiker-PRB-1988,
  Komiyama-PRB-1989, Chklovskii-1992, Haren-1995}, edge states (ESs) in the integer quantum Hall regime are still not fully understood
theoretically. In particular, despite intense activity~\cite{Ji2003, Marquardt-2004, Roulleau-2007,
  Litvin-2007, Roulleau-2008, Levkivskyi-2008,
  le-Sueur-Altimiras-PRL-2010,
  Altimiras-Nat-phys-2010,Altimiras-2010a,
  Lunde-2010,Otsuka-PRB-2010,Degiovanni-PRB-2010,Kovrizhin-Chalker-PRB-2011,Levkivskyi-Sukhorukov-PRB-2012,Karzig-Levchenko-Glazman-Oppen-NJP-2012,Chirolli-2013,Otsuka-et-al-2014, Inoue-2014,
  Helzel-2015,Tewari-2016,Slobodeniuk-2016,Gurman-2016}, our
understanding of the
dominant decoherence mechanism in transport through ESs is incomplete. This is illustrated by the recent experiment
of~\citet{Tewari-2016}, in which it was observed that decoherence of high energy electrons sent through a Mach Zehnder interferometer (MZI), formed with
two co-propagating ESs at filling
factor $\nu=2$, does not depend on the energy of
the injected electrons. This contradicts theoretical predictions based on
the Luttinger-liquid
model for one dimensional, translationally invariant
systems~\cite{Degiovanni-PRB-2010,Kovrizhin-Chalker-PRB-2011,Levkivskyi-Sukhorukov-PRB-2012,Tewari-2016,Slobodeniuk-2016}.
Disorder, however, is conspicuous for its absence in these approaches. While macroscopic phenomena, such as the quantization of
the Hall resistance, are robust to disorder, more subtle quantum effects, such
as coherent energy exchange and phase coherence between co-propagating ESs, can
be expected to be sensitive to even weak disorder at the edge of a
high mobility 2DEG~\cite{Pascher-2014}.

In this work, we show that by taking into account disorder, which
breaks translation invariance along the edge, a gapless continuum of
low energy quasi-particle excitations emerges. Their dynamics provides
a simple physical picture of
interaction-induced decoherence, which in turn provides a natural explanation for the experimental
findings of~\cite{Tewari-2016}. Our
theory has previously also been successfully applied to energy relaxation in
out-of-equilibrium ESs~\cite{Lunde-2010,
  Lunde-2016}.
In particular, in~\cite{Lunde-2016}, we showed that energy relaxation
of electrons injected high above the Fermi sea into the outermost of two
co-propagating, interacting and {\em weakly disordered} ESs,
can be described in terms of a drift-diffusion process of their energy
distribution function: As the injected electrons propagate along the outer
ES, they loose energy and their energy distribution moves towards the Fermi sea with a
{\em constant} energy drift velocity and broadens at a
position-dependent rate. The latter is
determined by the induced heating of the inner ES, which absorbs the
energy lost by the injected electrons and subsequently redistributes
part of this energy to the Fermi sea of the outer ES. The central new
idea of the present work is that, given a relation between the energy
and the phase of a propagating electron, knowing the dynamics of the
energy distribution function enables us to calculate the statistics of
the interaction-induced phase fluctuations. In the absence of
extrinsic dephasing mechanisms, the latter fully
determines the coherence of electron transport.
\begin{figure}[ht]
 \includegraphics[width=\columnwidth]{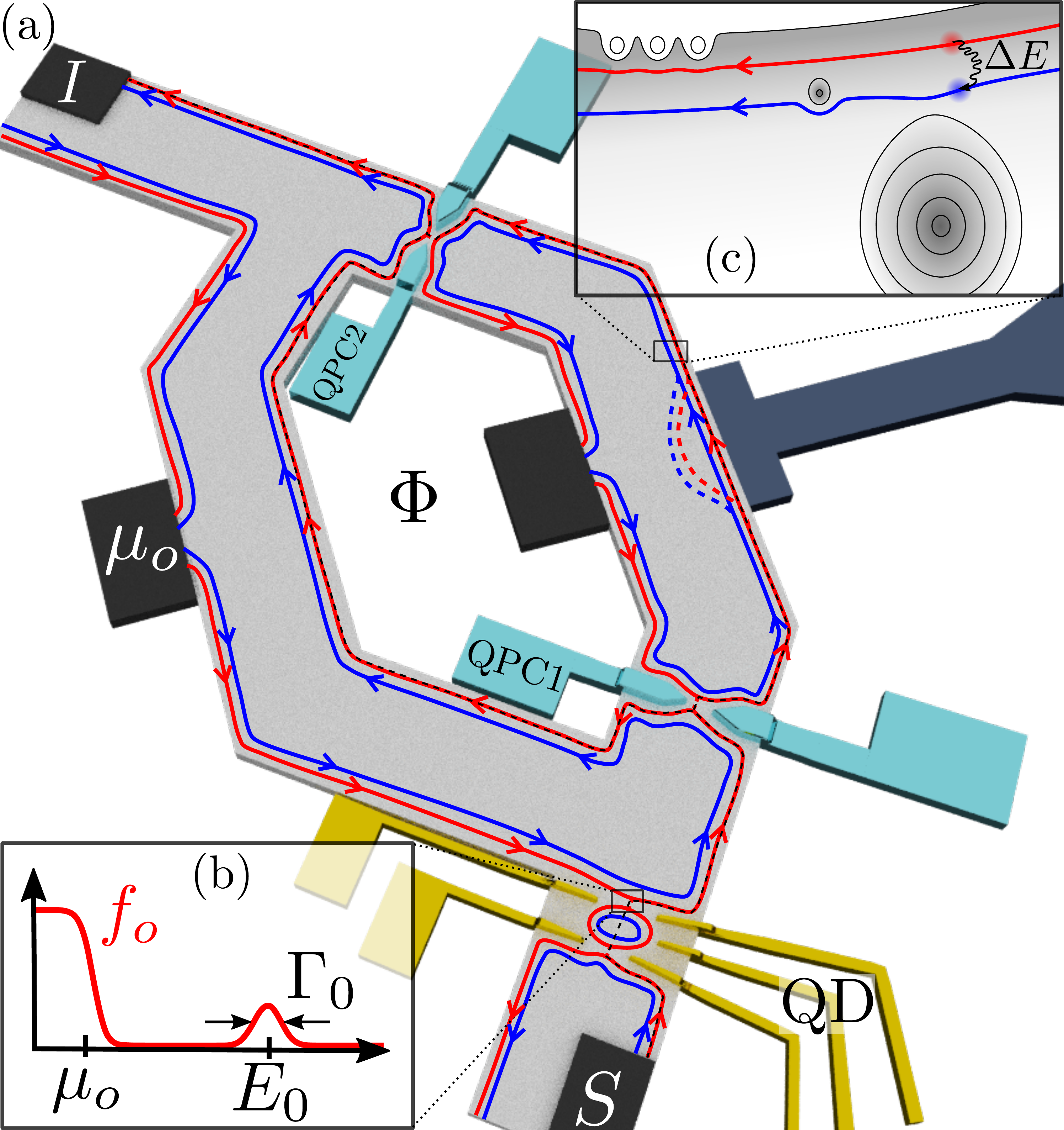}\caption{Schematics
   of a Mach-Zehnder interferomter realized in~\cite{Tewari-2016} with ESs of the
   integer quantum Hall effect at filling factor $\nu=2$. (a) The outer (inner)
   chiral ES is shown by a solid red (blue) line following the
   edge of the patterned 2DEG structure (light gray area). The arrows
   indicate the propagation direction which is determined by the
   orientation of the magnetic field perpendicular to the 2DEG (not
   shown). A quantum dot (bottom right) is used for
   energy-resolved injection of high-energy electrons, with mean
   energy $E_0$, into the
   outer ES by filtering the
   electrons emitted at the source marked with $S$. This creates a non-equilibrium
   energy distribution in the outer ES composed of a Fermi sea part
   and a narrow bump around $E_0$ as shown in (b). An injected electron is scattered at two QPCs and can
   follow two possible paths, marked by dashed (black) lines,
   before exiting the interferometer at the top left corner where the
   current is measured. Interference between the current amplitudes corresponding to
   these two paths can be modulated either by threading a magnetic
   flux $\Phi$ through the loop created by the two paths or, as in the
   experiment~\cite{Tewari-2016}, by applying a local gate voltage along one arm,
   in order to modify the path length difference $\Delta
   L$ (dark blue
   side gate). The
   visibility of the current oscillations in either $\Delta L$ or
   $\Phi$, is
   suppressed by inelastic scattering between electrons in the inner
   and outer ESs, which follow the equipotential lines of the
   disordered confinement potential~\cite{Pascher-2014} (c). In the case of unequal arm
   lengths, additional dephasing takes place due to the initial energy spread $\Gamma_0$ of the injected
   electrons (b).~\label{fig:MZI}}
\end{figure}

Explicitly, we find that the interaction-induced suppression of the
visibility of the current interference fringes through an electronic
MZI (see
Fig.~\ref{fig:MZI} (a)), is determined by the temperature of the electronic system, the drift velocity of the energy
distribution of the electrons injected into the outer ES and the heating of
the Fermi sea of the inner ES. Importantly, none of these quantities
depend on the injection energy of the electrons, resulting in
dephasing that is independent of the injection energy, inline with the
experiment of~\citet{Tewari-2016}. Rather, the amount of
dephasing is governed by the amount of disorder, which sets the available phase space for inelastic,
non-momentum conserving, electron-electron scattering.
This result suggests that disorder along the edges of a patterned 2DEG plays a more important role, with regards to energy
relaxation and decoherence of ESs, than hitherto assumed.

The system we consider is that of~\cite{Tewari-2016} and is depicted
  schematically in Fig.~\ref{fig:MZI}. It consists of two
  co-propagating chiral ESs, one of
  which is split via two quantum point contacts (QPCs) such as to form
  a MZI. Furthermore, a quantum dot (QD) side-coupled to the
  sample edge at the input of the interferometer is used for
  energy-resolved injection of electrons into the outer ES with an average energy $E_0$
  much larger than the Fermi energy $\mu_o$ of the outer
  ES as compared with the initial energy spread $\Gamma_0$ of the
  injected electrons, i.e. $E_0-\mu_o\gg\Gamma_0$ (see
  Fig.~\ref{fig:MZI} (b)). If
  all contacts, with the exception of the source, are kept at the
  same voltage, the DC current $I(\Phi,\Delta L)$
  measured at the output port of the interferometer will stem
  exclusively from electrons injected into the outer ES via the
  QD
  energy filter. The flux dependence of this
  current, is thus a sensitive
  probe of phase coherence along the outer ES. Instead of varying a
  magnetic flux $\Phi$ through the loop formed by the two arms of the
  interferometer, one may alternatively, as in the experiment~\cite{Tewari-2016},
  vary the path length difference $\Delta L=L_1-L_2$,
e.g. by applying a gate voltage to one of the arms. 
A similar
setup, albeit with a simpler topology, has been used previously to
investigate energy relaxation in out-of-equilibrium ESs~\cite{Altimiras-Nat-phys-2010,le-Sueur-Altimiras-PRL-2010}.

We focus on Coulomb mediated
energy exchange between the inner and outer co-propagating ESs,
without particle exchange. This is reasonable in the absence of
magnetic impurities, since the two edge
states have opposite spins and therefore particle exchange would
require a spin flip. 
In analogy with the
non-interacting scattering theory~\cite{Buttiker-1986,Blanter2001}, the contribution
from the outer ES to the current at the output port
of the MZI can then be written as
\begin{align}\label{eq:7}
I(\Phi,\Delta L)=\frac{e}{h}\int dEb_0(E)\braket{|r_1r_2+t_1t_2e^{i\phi_E}|^2}.
\end{align}
Here $b_0(E)$ denotes the energy distribution function of the
electrons injected via the QD
and centered at $E_0$. $r_i$ ($t_i$) is the real reflection (transmission)
probability amplitude at the $i$-th QPC, 
and $\phi_E$ denotes the relative phase acquired
by an electron injected with energy $E$ but traversing different arms of the MZI. Crucially, the phase $\phi_E$ is a random variable
that depends on the injection energy $E$, and on the random
energy exchange events between injection and detection. The brackets $\braket{\cdot}$ denote
averaging over all possible realizations of scattering
events. Assuming that the energy dependence of the transmission and reflection amplitudes
through the QPCs around the injection energy is negligible~\cite{Roulleau-2008,Tewari-2016}, it follows
from Eq.~(\ref{eq:7}) that the coherent part of the current is given
by
\begin{align}\label{eq:23}
I_{\varphi}=\frac{e}{h}(r_1r_2t_1t_2)\int
  dEb_0(E)\braket{e^{i\phi_E}+e^{-i\phi_E}}.
\end{align}
Hence, assuming $b_0(E)$ is known, our task is reduced to computing the average of $\exp(i\phi_E)$
over scattering events.

For the case of a linear dispersion considered
here, the phase acquired by an electron propagating in the outer ES along one of the
arms of the interferometer (say $l=1$ for the upper and $l=2$ for
the lower arm according to Fig.~\ref{fig:MZI}) is simply given by
\begin{align}\label{eq:9}
\phi_E^{(l)}(x)=\frac{1}{\hbar v_o}\int_0^{x}dyE(y),
\end{align}
where $v_o$ is the velocity of
the electron in the outer ES and $E(y)$ denotes the energy of the electron at position $y$ in arm $l$,
given the initial energy $E(0)=E$. The relative phase at the detector is then simply
given by
\begin{align}\label{eq:24}
\phi_E=\phi_E^{(1)}(L_1)-\phi_E^{(2)}(L_2) + 2\pi\Phi/\Phi_0,
\end{align}
where $L_l$ is the length of
interferometer arm $l$, possibly including a gate induced path length
variation. Because electrons on different arms
do not interact, owing to screening and a sufficiently large spatial separation, the average over scattering events factorizes
\begin{align}\label{eq:25}
\braket{e^{i\phi_E}}=\braket{\exp(i\phi_E^{(1)}(L_1))}\braket{\exp(-i\phi_E^{(2)}(L_2))}e^{2\pi
i\Phi/\Phi_0},
\end{align}
and it is sufficient to evaluate the interaction-induced
coherence suppression factor
$\mathcal{F}_E^{(l)}(x)\equiv\braket{\exp(i\phi_E^{(l)}(x))}$ for one arm.
From now on, we thus 
suppress the arm label $l$.

Our starting point for evaluating $\mathcal{F}_E(x)$, is the kinetic
Boltzmann equation for the
energy distribution functions $f_\alpha$ of the inner ($\alpha=i$) and outer
($\alpha=o$) ESs
\begin{align}\label{eq:6}
v_{\alpha}\partial_xf_\alpha(E,x)=\mathcal{I}_{Ex\alpha}\left[ f_{\alpha},f_{\bar\alpha} \right].
\end{align}
The term on the right-hand side is the difference of
in-scattering ($\{E'\}\rightarrow E$) and
out-scattering ($E\rightarrow \{E'\}$) energy exchange processes
between the inner and outer ESs~\cite{footnote-1}. Here and below, we
use the shorthand notation $\overline\alpha=\delta_{\alpha i}o
+\delta_{\alpha o} i$. If both energy and momentum are conserved, then two-body
collisions cannot change the distribution function in one dimension, as long as $v_i\not=
v_o$,~\cite{footnote-2}. However,
disorder along the edge breaks translation invariance such that
inelastic electron-electron scattering without momentum conservation
becomes possible. Thereby, an effective interaction is induced and the
phase-space for energy exchange between electrons in the inner and
outer ESs opens up. In contrast to collective excitations in a finite
length system~\cite{Bocquillon-2013}, these excitations are gapless. As shown
in~\cite{supp_mat,Lunde-2010,Lunde-2016,supp_mat}, this situation is
described by Eq.~(\ref{eq:6}) with the
collision integral
\begin{align}\label{eq:3}
\mathcal{I}_{Ex\alpha}[f_{\alpha},f_{\bar\alpha}]&=v_{\alpha}\gamma\int d\omega
  e^{-(\omega/\Delta E)^2}\nonumber\\
&\times\Big\{
  f_{\alpha}(E+\omega,x)\left[1-f_{\alpha}(E,x)\right]D_{\bar\alpha}(\omega,x)\nonumber\\
&-f_{\alpha}(E,x)\left[1-f_{\alpha}(E+\omega,x)\right]D_{\bar\alpha}(-\omega,x) \Big\},
\end{align}
where $\gamma$ is the effective inter-ES interaction strength, $\Delta
E$ is the energy scale for the amount of energy exchanged per
non-momentum conserving collision~\cite{supp_mat}, and
\begin{align}\label{eq:10}
D_\alpha(\omega,x)=\int dE f_\alpha(E-\omega,x)(1-f_\alpha(E,x)).
\end{align}
The inner ES is initially in thermal equilibrium so that
$f_i(E,0)=1/[1+\exp((E-\mu_i)/\kb T)]$. Furthermore, because we consider electrons injected high above
the Fermi sea 
in the outer ES ($E_0-\mu_o\gg\Gamma_0$), we can split the distribution function in the outer ES
into two essentially non-overlapping contributions
\begin{align}
f_o(E,x)=\fo(E,x) + b(E,x),
\end{align}
where $\fo(E,x=0)=1/[1+\exp(\beta (E-\mu_o))]$ and $b(E,x)$ is the
energy distribution of the injected electrons at position $x$ with
boundary condition $b(E,0)=b_0(E)$. If,
as in the experiment~\cite{Tewari-2016}, the transmission
probability through the QD is small, then $b(E,x)\ll 1$. Consequently, we can
neglect, in the collision integral, all terms of order
$\mathcal{O}(b^2,b\fo)$. Finally, since we are interested in the limit of weak disorder, $\Delta E$ is taken to be the smallest energy
scale, e.g. $\Delta E\ll \kb T,\Gamma_0$. These steps allow us to
derive, from the kinetic equation, the following set of coupled Fokker-Planck
equations~\cite{Lunde-2016}
\begin{subequations}
\label{FPeqs}
\begin{align}
\partial_xb(E,x)&=\eta\left\{ \partial_Eb(E,x)+D_i(0,x)\partial_E^2b(E,x) \right\}\label{eq:8},\\
\partial_xf_i(E,x)&=\eta\frac{N_b}{\rho^{}_o}\partial_E^2f_i(E,x)\label{eq:4}\\
&+\eta\left\{ \left[ 1-2f_i(E,x)
  \right]\partial_Ef_i(E,x)+\frak{D}_o(0,x)\partial_E^2f_i(E,x)
  \right\},\nonumber\\
\partial_x\fo(E,x)&=\eta\left\{ \left[ 1-2\fo(E,x) \right]\partial_E\fo(E,x)+D_i(0,x)\partial_E^2\fo(E,x) \right\}\label{eq:15}.
\end{align}
\end{subequations}
Here $\eta=(\sqrt{\pi}/4)\gamma(\Delta E)^3$ is the energy drift
velocity, $\rho^{}_o$ is the density of states in
the outer ES, $N_b=\rho_o\int dEb_0(E)$, is the mean number of
injected electrons and $\frak{D}_o(\omega,x)=\int dE \fo(E-\omega,x)\left[ 1-\fo(E,x)
\right]$.

The Fokker-Planck equation~(\ref{eq:8}) is equivalent~\cite{Gardiner} to the It\^o
stochastic differential equation
\begin{align}\label{eq:1}
d E = -\eta d x + g(x)\dd W_x,
\end{align}
where $g(x)=\sqrt{2\eta D_i(0,x)}$ and $\dd W_x$ is a Wiener
process. The random energy of an electron injected at $x=0$ with
energy $E$, is
obtained by integrating Eq.~(\ref{eq:1}) and using the initial condition
$E(x=0)=E$:
\begin{align}\label{eq:2}
E(x) = E-\eta x +\int_0^xg(y)\dd W_y.
\end{align}
The last term in Eq.~(\ref{eq:2}) is a stochastic It\^o integral. By
applying the It\^o calculus
($\braket{\dd W_x\dd W_{x'}}=\delta(x-x')d x$, $\braket{\dd
W_x}=0$), we find the mean and variance (${\rm
Var}[\cdot]=\braket{(\cdot)^2}-\braket{\cdot}^2$) of the energy at
position $x$ as
\begin{subequations}
\begin{align}
\braket{E(x)}=E-\eta x,\label{eq:11}\\
{\rm Var}[E(x)]=2\eta\int_0^xD_i(0,y)d y.\label{eq:16}
\end{align}
\end{subequations}
Note that averaging Eq.~(\ref{eq:11}) over the injection
energy using the probability density $(\rho^{}_o/N_b)b_0(E)$ yields $\braket{\braket{E(x)}}_0=E_0-\eta x$,
which explains why $\eta$ is called the energy drift
velocity of the energy distribution of the injected electrons. Because of
Eq.~(\ref{eq:16}), we further call $2\eta D_i(0,x)$ the dynamic
diffusion coefficient~\cite{Lunde-2016}. According to Eq.~(\ref{eq:9}), the phase of the electron at position
$x$ is now given by
\begin{align}
\phi_E(x) &=\frac{1}{\hbar v_o}\left(
  Ex-\frac{1}{2}\eta x^2 \right)+\frac{1}{\hbar
  v_o}\int_0^x\int_0^yg(z)\dd W_zd y.
\end{align}
Using again the It\^o calculus~\cite{supp_mat}, the last integral can be rewritten as
\begin{align}
\int_0^x\int_0^yg(z)\dd W_zd y=\int_0^x(x-y)
  g(y)\dd W_y,
\end{align} 
from which it follows that the variance of the phase is
\begin{align}\label{eq:20}
\delta\phi^2(x)\equiv{\rm Var}\left[ \phi_E(x) \right]&=\frac{2\eta}{\left( \hbar v_o \right)^2}\int_0^x(x-y)^2D_i(0,y)d y.
\end{align}
Because the fluctuating part of the phase is itself a Gaussian
random variable with zero mean, we can use the identity
$\braket{\exp(i\phi)}=\exp(i\braket{\phi})\exp(-\delta\phi^2/2)$, and the interaction-induced dephasing factor is given by
\begin{align}\label{eq:5}
\mathcal{F}_E(x)&=\exp\left( \frac{i}{\hbar v_o}\left[
                  Ex-\frac{1}{2}\eta x^2 \right] \right)\exp\left(
                  -\frac{\delta\phi^2(x)}{2}
  \right).
\end{align}
Eq.~(\ref{eq:5}) together with Eq.~(\ref{eq:20}) are the main analytic results of this work. They link
the interaction-induced phase coherence suppression factor of the outer ES
to the relaxation induced smearing of the energy distribution of the
inner ES, quantified by $D_i(0,x)$ (see Eq.~(\ref{eq:10})). Importantly, the latter is independent of the injection
energy as shown below. Combining
Eqs.~(\ref{eq:23}),~(\ref{eq:24}),~(\ref{eq:25}) and~(\ref{eq:5}), we obtain an explicit
expression for the coherent current through the
interferometer
\begin{align}\label{eq:21}
I_{\varphi}(\Phi,\Delta
  L)&=\frac{2e}{h}\frac{N_b}{\rho_o^{}}(r_1r_2t_1t_2)B_0(\Delta L)e^{-\frac{1}{2}\left[
                  \delta\phi^2(L_1)+\delta\phi^2(L_2) \right]}\\
&\times\cos\left(\frac{E_0 \Delta L }{\hbar v_o} -\frac{\eta\left( L_1^2-L_2^2 \right)}{2\hbar v_o}+\frac{2\pi\Phi}{\Phi_0}\right)\nonumber.
\end{align}
Here the factor
$B_0(\Delta L)=\frac{\rho_o^{}}{N_b}\int dE b_0(E+E_0)e^{i\frac{E\Delta
    L}{\hbar v_o}}$ characterizes the dephasing due to the initial
energy spread of the injected electrons for finite
path length difference, and the
exponential factor quantifies the interaction-induced
dephasing. For an initial Gaussian energy distribution of the form
$b_0(E)=\frac{N_b}{\rho_o^{}\sqrt{\pi}\Gamma_0}\exp\left[ (E-E_0)^2/\Gamma_0^2
\right]$, we have $B_0(\Delta L)=\exp\left[ -\left( \frac{\Gamma_0\Delta
    L}{2\hbar v_o} \right)^2 \right]$. For an initial distribution of
the form $b_0(E)=\frac{N_b}{4\Gamma_0\rho_o^{}}\cosh^{-2}\left(
  \frac{E-E_0}{2\Gamma_0} \right)$, which, with $\Gamma_0=\kb T$, is
appropriate for injection through a
thermally broadened QD level~\cite{Tewari-2016}, we have
$B_0(\Delta L)=\frac{\pi\Gamma_0\Delta L}{2\hbar
  v_o}{\rm csch}\left( \frac{\pi\Gamma_0\Delta L}{\hbar v_o}
\right)$. In both cases $B_0(\Delta L\rightarrow 0)\rightarrow 1$ as expected.
The experimentally relevant visibility
of the current interference $\mathcal{V}\equiv(I_{\rm MAX}-I_{\rm MIN})/(I_{\rm
  MAX}+I_{\rm MIN})$
is found by extremizing the cosine in Eq.~(\ref{eq:21}) over either $\Phi$ or variations of
the path length difference $\Delta L$~\cite{footnote-4} and reads
\begin{align}\label{eq:22}
\mathcal{V}&=\frac{2r_1r_2t_1t_2}{(r_1r_2)^2+(t_1t_2)^2}B_0(\Delta L)e^{-\frac{1}{2}\left[ \delta\phi^2(L_1)+\delta\phi^2(L_2) \right]}.
\end{align}

To obtain the variance of the phase fluctuations, we
need to evaluate the function $D_i(0,x)$, i.e. solve
Eqs.~(\ref{eq:4}) and~(\ref{eq:15}). A thorough discussion of these equations can be found
in~\cite{Lunde-2016}, where it was shown that an approximate
solution, takes the form of an effective
temperature ansatz for $f_i(E,x)$ and $\frak{f}_o(E,x)$:
\begin{subequations}
\begin{align}
f^F_i(E,x)&=\frac{1}{1+\exp\left[ \frac{E-\mu_i}{\kb T_i(x)}
            \right]},\label{eq:17}\\
\fo^F(E,x)&=\frac{1}{1+\exp\left[ \frac{E-\mu_o}{\kb T_o(x)} \right]}.\label{eq:18}
\end{align}
\end{subequations}
From Eq.~(\ref{eq:10}) it immediately follows that, within the effective
temperature approximation, $D_i(0,x)= \kb T_i(x)$.
The coupled Fokker-Planck Eqs.~(\ref{eq:4}) and~(\ref{eq:15}) now reduce to coupled
ordinary differential equations for $T_i(x)$
and $T_o(x)$:
\begin{subequations}
\label{eq:19}
\begin{align}
\kb \partial_xT_i(x)&=\eta\frac{3}{\pi^2}\left( \frac{N_b/\rho^{}_o}{\kb
                      T_i(x)}+\frac{T_o(x)}{T_i(x)}-1 \right),\label{eq:12}\\
\kb \partial_xT_o(x)&=\eta\frac{3}{\pi^2}\left(\frac{T_i(x)}{T_o(x)}-1 \right).\label{eq:14}
\end{align}
\end{subequations}
In the case of interest here, the Fermi seas of the inner and outer ESs
initially have the same temperature $T_i(0)=T_o(0)=T$. Moreover, since
we are working in the limit of weak disorder where $\Delta E\ll\kb T$, it is
reasonable to expect that the difference between the two effective
temperatures $T_d(x)=(T_i(x)-T_o(x))/2$ remains small compared
with the sum of the temperatures $T_s(x)=(T_i(x)+T_o(x))/2$ at all
positions. From this assumption, one can then
derive an approximate solution of~(\ref{eq:19}) which yields~\cite{Lunde-2016}
\begin{align}\label{eq:26}
\kb T_i(x)&\simeq\kb T\sqrt{1+\frac{x}{x_s}}+\frac{N_b}{4\rho^{}_o}\left( 1-e^{\frac{4\kb
            T}{N_b/\rho^{}_o}\left[  1-\sqrt{1+\frac{x}{x_s}} \right]} \right),
\end{align}
with $x_s=(\pi\kb T)^2\rho_o^{}/(3\eta N_b)$. At short distances $x\ll x_s$, we have $T_i(x)\simeq
T\left( 1+x/x_s \right)$, in which case the integral in~(\ref{eq:20}) can be evaluated
analytically, yielding
\begin{align}\label{eq:27}
\left|\mathcal{F}_E(x)\right|^2
  \simeq\exp\left[ -\frac{2}{3}\frac{\eta
  k_BT}{(\hbar v_o)^2}\left(x^3+\frac{x^4}{12x_s}\right)\right],\quad\text{for}\, x\ll x_s.
\end{align}
Hence, the smaller the propagation velocity $v_o$, the stronger the
dephasing, a trend which was recently observed by~\citet{Gurman-2016}.
At large distances $x\gg x_s$, the exponential term in Eq.~(\ref{eq:26})
vanishes and $T_i(x)\simeq\kb T\sqrt{1+x/x_s}+N_b/(4\rho^{}_o)$.

\begin{figure}[ht]
 \includegraphics[width=\columnwidth]{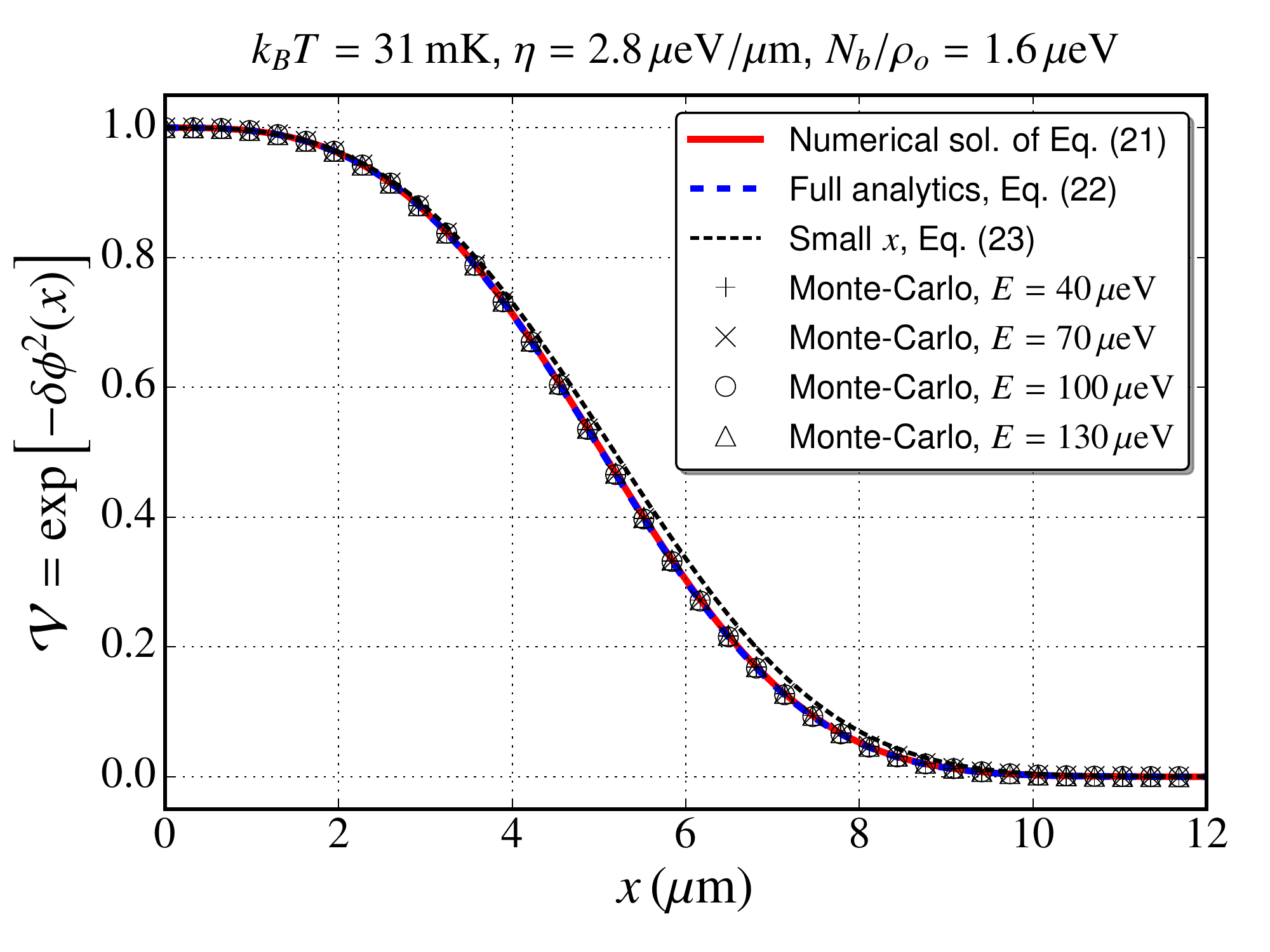}\caption{Visibility
   of the current interference fringes (\ref{eq:22}) as a function of the interferometer
   arm length for $\Delta L=0$, $r_1=r_2=t_1=t_2=1/\sqrt{2}$, and
   $v_o=5\cdot 10^4\,{\rm m/s}$. The
   solid (red) curve shows the result
   obtained by numerically integrating the
   differential equations~(\ref{eq:19}). The dashed (blue) curve
   shows the analytic result obtained using Eq.~(\ref{eq:26}). The thin
   dashed (black) curve shows Eq.~(\ref{eq:27}), obtained from the short
   distance limit of~(\ref{eq:26}), when $x\ll x_s\approx 5.2\,{\rm
     \mu m}$. The symbols show results from Monte Carlo
   simulations of the kinetic equation for different injection energies
   $E\in\{40,70,100,130\}\,{\rm\mu eV}$~\cite{supp_mat}.\label{fig:visibility}}
\end{figure}

From the data in~\cite{Tewari-2016}, we can estimate that in the experiment,
$N_b/\rho_o^{}\approx 1.6\,{\rm \mu eV}$~\cite{supp_mat}. The only remaining free parameter $\eta$ can then be
determined by fitting Eq.~(\ref{eq:22}) to the measured visibility, using the
experimentally determined values for the other parameters~\cite{Tewari-2016}: $\kb
T\approx 31\,{\rm mK}$, $v_o\approx 5\cdot 10^{4}{\rm ms^{-1}}$ and
$r_1=r_2=t_1=t_2\approx 1/\sqrt{2}$, as
well as $L_1=L_2=L\approx 7.2\,{\rm \mu m}$. This yields~\cite{supp_mat} an energy drift velocity of $\eta\approx
2.8\,{\rm \mu eV}/{\rm \mu m}$.
This value further justifies our perturbative analysis for weak momentum conservation
breaking of
the experiment~\cite{Tewari-2016}, where a visibility independent of
the injection energy is observed in the range $
(E_0-\mu_o)\in[30,130]\,{\rm \mu eV}> \eta L\approx 28\,{\mu
  eV}$. The regime where $\eta L > E_0-\mu_o$ is outside of the
Fokker-Planck regime, since the distribution in the outer ES can no
longer be separated into two non-overlapping
contributions. Experimentally, dephasing in this regime is observed to
depend on the injection energy~\cite{Tewari-2016}.
Using the above estimates, we plot the
visibility according to Eq.~(\ref{eq:22}), with $\Delta L=0$, as a function of the
interferometer length in Fig.~\ref{fig:visibility}.

To further validate
our analytic results, we compare them with the results from a Monte Carlo simulation of the
Fokker-Planck dynamics of the kinetic equation, for
different values of the injection energy. In this simulation,
we discretize the stochastic energy exchange process for a
given injection energy. At each step, we
determine the scattering rate and the distribution of scattering
energies from Eq.~(\ref{eq:3}). We then use these to update the energy and
accumulated phase of an
electron as it propagates along the edge. The phase suppression factor
is estimated by averaging over many such ``trajectories'':
$\mathcal{F}_E(x)=(1/M)\sum_{m=1}^M\exp(i\phi_{E,m}(x))$. Further details on our implementation are
given in~\cite{supp_mat}. The results confirm our
analytic predictions (see
Fig.~\ref{fig:visibility}).

In conclusion, we have shown how the interplay of disorder and Coulomb
interaction leads to the loss of phase
coherence of the current through a MZI formed with two
co-propagating ESs of the integer quantum Hall effect. Crucially we
find that dephasing does not depend on the injection energy, in
agreement with recent
experiments~\cite{Tewari-2016}. Furthermore, our theory makes
quantitative predictions for the length dependence of the dephasing
(see Fig.~\ref{fig:visibility} and Eqs.~(\ref{eq:20}),~(\ref{eq:5}) and~(\ref{eq:26})), which could easily be tested
by adapting existing experimental systems.

{\em Acknowledgments}. We thank Karsten Flensberg for discussion. The
Monte Carlo simulations were performed in
a parallel computing environment at sciCORE
(\url{http://scicore.unibas.ch/}) scientific computing core facility
at University of Basel.
SEN acknowledges financial support from the
Swiss NSF and AML from the Carlsberg Foundation.

\bibliography{Refs}

\begin{thebibliography}{47}
\expandafter\ifx\csname natexlab\endcsname\relax\def\natexlab#1{#1}\fi
\expandafter\ifx\csname bibnamefont\endcsname\relax
  \def\bibnamefont#1{#1}\fi
\expandafter\ifx\csname bibfnamefont\endcsname\relax
  \def\bibfnamefont#1{#1}\fi
\expandafter\ifx\csname citenamefont\endcsname\relax
  \def\citenamefont#1{#1}\fi
\expandafter\ifx\csname url\endcsname\relax
  \def\url#1{\texttt{#1}}\fi
\expandafter\ifx\csname urlprefix\endcsname\relax\def\urlprefix{URL }\fi
\providecommand{\bibinfo}[2]{#2}
\providecommand{\eprint}[2][]{\url{#2}}

\bibitem[{\citenamefont{Tsui et~al.}(1982)\citenamefont{Tsui, Stormer, and
  Gossard}}]{Tsui-1982}
\bibinfo{author}{\bibfnamefont{D.~C.} \bibnamefont{Tsui}},
  \bibinfo{author}{\bibfnamefont{H.~L.} \bibnamefont{Stormer}},
  \bibnamefont{and} \bibinfo{author}{\bibfnamefont{A.~C.}
  \bibnamefont{Gossard}}, \bibinfo{journal}{Phys. Rev. Lett.}
  \textbf{\bibinfo{volume}{48}}, \bibinfo{pages}{1559} (\bibinfo{year}{1982}).

\bibitem[{\citenamefont{Wen}(1990)}]{Wen-1990}
\bibinfo{author}{\bibfnamefont{X.~G.} \bibnamefont{Wen}},
  \bibinfo{journal}{Phys. Rev. B} \textbf{\bibinfo{volume}{41}},
  \bibinfo{pages}{12838} (\bibinfo{year}{1990}).

\bibitem[{\citenamefont{Das~Sarma et~al.}(2005)\citenamefont{Das~Sarma,
  Freedman, and Nayak}}]{Sarma-2005}
\bibinfo{author}{\bibfnamefont{S.}~\bibnamefont{Das~Sarma}},
  \bibinfo{author}{\bibfnamefont{M.}~\bibnamefont{Freedman}}, \bibnamefont{and}
  \bibinfo{author}{\bibfnamefont{C.}~\bibnamefont{Nayak}},
  \bibinfo{journal}{Phys. Rev. Lett.} \textbf{\bibinfo{volume}{94}},
  \bibinfo{pages}{166802} (\bibinfo{year}{2005}).

\bibitem[{\citenamefont{Huynh et~al.}(2012)\citenamefont{Huynh, Portier,
  le~Sueur, Faini, Gennser, Mailly, Pierre, Wegscheider, and
  Roche}}]{Huynh-2012}
\bibinfo{author}{\bibfnamefont{P.-A.} \bibnamefont{Huynh}},
  \bibinfo{author}{\bibfnamefont{F.}~\bibnamefont{Portier}},
  \bibinfo{author}{\bibfnamefont{H.}~\bibnamefont{le~Sueur}},
  \bibinfo{author}{\bibfnamefont{G.}~\bibnamefont{Faini}},
  \bibinfo{author}{\bibfnamefont{U.}~\bibnamefont{Gennser}},
  \bibinfo{author}{\bibfnamefont{D.}~\bibnamefont{Mailly}},
  \bibinfo{author}{\bibfnamefont{F.}~\bibnamefont{Pierre}},
  \bibinfo{author}{\bibfnamefont{W.}~\bibnamefont{Wegscheider}},
  \bibnamefont{and} \bibinfo{author}{\bibfnamefont{P.}~\bibnamefont{Roche}},
  \bibinfo{journal}{Phys. Rev. Lett.} \textbf{\bibinfo{volume}{108}},
  \bibinfo{pages}{256802} (\bibinfo{year}{2012}).

\bibitem[{\citenamefont{Bocquillon et~al.}(2014)\citenamefont{Bocquillon,
  Freulon, Parmentier, Berroir, Pla\c{c}ais, Wahl, Rech, Jonckheere, Martin,
  Grenier et~al.}}]{Bocquillon-2014}
\bibinfo{author}{\bibfnamefont{E.}~\bibnamefont{Bocquillon}},
  \bibinfo{author}{\bibfnamefont{V.}~\bibnamefont{Freulon}},
  \bibinfo{author}{\bibfnamefont{F.~D.} \bibnamefont{Parmentier}},
  \bibinfo{author}{\bibfnamefont{J.-M.} \bibnamefont{Berroir}},
  \bibinfo{author}{\bibfnamefont{B.}~\bibnamefont{Pla\c{c}ais}},
  \bibinfo{author}{\bibfnamefont{C.}~\bibnamefont{Wahl}},
  \bibinfo{author}{\bibfnamefont{J.}~\bibnamefont{Rech}},
  \bibinfo{author}{\bibfnamefont{T.}~\bibnamefont{Jonckheere}},
  \bibinfo{author}{\bibfnamefont{T.}~\bibnamefont{Martin}},
  \bibinfo{author}{\bibfnamefont{C.}~\bibnamefont{Grenier}},
  \bibnamefont{et~al.}, \bibinfo{journal}{Annalen der Physik}
  \textbf{\bibinfo{volume}{526}}, \bibinfo{pages}{1} (\bibinfo{year}{2014}).

\bibitem[{\citenamefont{Haack et~al.}(2014)\citenamefont{Haack, Albert, and
  Flindt}}]{Haack-2014}
\bibinfo{author}{\bibfnamefont{G.}~\bibnamefont{Haack}},
  \bibinfo{author}{\bibfnamefont{M.}~\bibnamefont{Albert}}, \bibnamefont{and}
  \bibinfo{author}{\bibfnamefont{C.}~\bibnamefont{Flindt}},
  \bibinfo{journal}{Phys. Rev. B} \textbf{\bibinfo{volume}{90}},
  \bibinfo{pages}{205429} (\bibinfo{year}{2014}).

\bibitem[{\citenamefont{Hofer and Sothmann}(2015)}]{Hofer-2015}
\bibinfo{author}{\bibfnamefont{P.~P.} \bibnamefont{Hofer}} \bibnamefont{and}
  \bibinfo{author}{\bibfnamefont{B.}~\bibnamefont{Sothmann}},
  \bibinfo{journal}{Phys. Rev. B} \textbf{\bibinfo{volume}{91}},
  \bibinfo{pages}{195406} (\bibinfo{year}{2015}).

\bibitem[{\citenamefont{S\'anchez et~al.}(2015)\citenamefont{S\'anchez,
  Sothmann, and Jordan}}]{Sanchez-2015}
\bibinfo{author}{\bibfnamefont{R.}~\bibnamefont{S\'anchez}},
  \bibinfo{author}{\bibfnamefont{B.}~\bibnamefont{Sothmann}}, \bibnamefont{and}
  \bibinfo{author}{\bibfnamefont{A.~N.} \bibnamefont{Jordan}},
  \bibinfo{journal}{Phys. Rev. Lett.} \textbf{\bibinfo{volume}{114}},
  \bibinfo{pages}{146801} (\bibinfo{year}{2015}).

\bibitem[{\citenamefont{Halperin}(1982)}]{Halperin-PRB-1982}
\bibinfo{author}{\bibfnamefont{B.~I.} \bibnamefont{Halperin}},
  \bibinfo{journal}{Phys. Rev. B} \textbf{\bibinfo{volume}{25}},
  \bibinfo{pages}{2185} (\bibinfo{year}{1982}).

\bibitem[{\citenamefont{B{\"u}ttiker}(1988)}]{Buttiker-PRB-1988}
\bibinfo{author}{\bibfnamefont{M.}~\bibnamefont{B{\"u}ttiker}},
  \bibinfo{journal}{Phys. Rev. B} \textbf{\bibinfo{volume}{38}},
  \bibinfo{pages}{9375} (\bibinfo{year}{1988}).

\bibitem[{\citenamefont{Komiyama et~al.}(1989)\citenamefont{Komiyama, Hirai,
  Sasa, and Hiyamizu}}]{Komiyama-PRB-1989}
\bibinfo{author}{\bibfnamefont{S.}~\bibnamefont{Komiyama}},
  \bibinfo{author}{\bibfnamefont{H.}~\bibnamefont{Hirai}},
  \bibinfo{author}{\bibfnamefont{S.}~\bibnamefont{Sasa}}, \bibnamefont{and}
  \bibinfo{author}{\bibfnamefont{S.}~\bibnamefont{Hiyamizu}},
  \bibinfo{journal}{Phys. Rev. B} \textbf{\bibinfo{volume}{40}},
  \bibinfo{pages}{12566} (\bibinfo{year}{1989}).

\bibitem[{\citenamefont{Chklovskii et~al.}(1992)\citenamefont{Chklovskii,
  Shklovskii, and Glazman}}]{Chklovskii-1992}
\bibinfo{author}{\bibfnamefont{D.~B.} \bibnamefont{Chklovskii}},
  \bibinfo{author}{\bibfnamefont{B.~I.} \bibnamefont{Shklovskii}},
  \bibnamefont{and} \bibinfo{author}{\bibfnamefont{L.~I.}
  \bibnamefont{Glazman}}, \bibinfo{journal}{Phys. Rev. B}
  \textbf{\bibinfo{volume}{46}}, \bibinfo{pages}{4026} (\bibinfo{year}{1992}).

\bibitem[{\citenamefont{van Haren et~al.}(1995)\citenamefont{van Haren, Blom,
  and Wolter}}]{Haren-1995}
\bibinfo{author}{\bibfnamefont{R.~J.~F.} \bibnamefont{van Haren}},
  \bibinfo{author}{\bibfnamefont{F.~A.~P.} \bibnamefont{Blom}},
  \bibnamefont{and} \bibinfo{author}{\bibfnamefont{J.~H.}
  \bibnamefont{Wolter}}, \bibinfo{journal}{Phys. Rev. Lett.}
  \textbf{\bibinfo{volume}{74}}, \bibinfo{pages}{1198} (\bibinfo{year}{1995}).

\bibitem[{\citenamefont{Ji et~al.}(2003)\citenamefont{Ji, Chung, Sprinzak,
  Heiblum, Mahalu, and Shtrikman}}]{Ji2003}
\bibinfo{author}{\bibfnamefont{Y.}~\bibnamefont{Ji}},
  \bibinfo{author}{\bibfnamefont{Y.}~\bibnamefont{Chung}},
  \bibinfo{author}{\bibfnamefont{D.}~\bibnamefont{Sprinzak}},
  \bibinfo{author}{\bibfnamefont{M.}~\bibnamefont{Heiblum}},
  \bibinfo{author}{\bibfnamefont{D.}~\bibnamefont{Mahalu}}, \bibnamefont{and}
  \bibinfo{author}{\bibfnamefont{H.}~\bibnamefont{Shtrikman}},
  \bibinfo{journal}{Nature} \textbf{\bibinfo{volume}{422}},
  \bibinfo{pages}{415} (\bibinfo{year}{2003}).

\bibitem[{\citenamefont{Marquardt and Bruder}(2004)}]{Marquardt-2004}
\bibinfo{author}{\bibfnamefont{F.}~\bibnamefont{Marquardt}} \bibnamefont{and}
  \bibinfo{author}{\bibfnamefont{C.}~\bibnamefont{Bruder}},
  \bibinfo{journal}{Phys. Rev. Lett.} \textbf{\bibinfo{volume}{92}},
  \bibinfo{pages}{056805} (\bibinfo{year}{2004}).

\bibitem[{\citenamefont{Roulleau et~al.}(2007)\citenamefont{Roulleau, Portier,
  Glattli, Roche, Cavanna, Faini, Gennser, and Mailly}}]{Roulleau-2007}
\bibinfo{author}{\bibfnamefont{P.}~\bibnamefont{Roulleau}},
  \bibinfo{author}{\bibfnamefont{F.}~\bibnamefont{Portier}},
  \bibinfo{author}{\bibfnamefont{D.~C.} \bibnamefont{Glattli}},
  \bibinfo{author}{\bibfnamefont{P.}~\bibnamefont{Roche}},
  \bibinfo{author}{\bibfnamefont{A.}~\bibnamefont{Cavanna}},
  \bibinfo{author}{\bibfnamefont{G.}~\bibnamefont{Faini}},
  \bibinfo{author}{\bibfnamefont{U.}~\bibnamefont{Gennser}}, \bibnamefont{and}
  \bibinfo{author}{\bibfnamefont{D.}~\bibnamefont{Mailly}},
  \bibinfo{journal}{Phys. Rev. B} \textbf{\bibinfo{volume}{76}},
  \bibinfo{pages}{161309} (\bibinfo{year}{2007}).

\bibitem[{\citenamefont{Litvin et~al.}(2007)\citenamefont{Litvin, Tranitz,
  Wegscheider, and Strunk}}]{Litvin-2007}
\bibinfo{author}{\bibfnamefont{L.~V.} \bibnamefont{Litvin}},
  \bibinfo{author}{\bibfnamefont{H.-P.} \bibnamefont{Tranitz}},
  \bibinfo{author}{\bibfnamefont{W.}~\bibnamefont{Wegscheider}},
  \bibnamefont{and} \bibinfo{author}{\bibfnamefont{C.}~\bibnamefont{Strunk}},
  \bibinfo{journal}{Phys. Rev. B} \textbf{\bibinfo{volume}{75}},
  \bibinfo{pages}{033315} (\bibinfo{year}{2007}).

\bibitem[{\citenamefont{Roulleau et~al.}(2008)\citenamefont{Roulleau, Portier,
  Roche, Cavanna, Faini, Gennser, and Mailly}}]{Roulleau-2008}
\bibinfo{author}{\bibfnamefont{P.}~\bibnamefont{Roulleau}},
  \bibinfo{author}{\bibfnamefont{F.}~\bibnamefont{Portier}},
  \bibinfo{author}{\bibfnamefont{P.}~\bibnamefont{Roche}},
  \bibinfo{author}{\bibfnamefont{A.}~\bibnamefont{Cavanna}},
  \bibinfo{author}{\bibfnamefont{G.}~\bibnamefont{Faini}},
  \bibinfo{author}{\bibfnamefont{U.}~\bibnamefont{Gennser}}, \bibnamefont{and}
  \bibinfo{author}{\bibfnamefont{D.}~\bibnamefont{Mailly}},
  \bibinfo{journal}{Phys. Rev. Lett.} \textbf{\bibinfo{volume}{101}},
  \bibinfo{pages}{186803} (\bibinfo{year}{2008}).

\bibitem[{\citenamefont{Levkivskyi and Sukhorukov}(2008)}]{Levkivskyi-2008}
\bibinfo{author}{\bibfnamefont{I.~P.} \bibnamefont{Levkivskyi}}
  \bibnamefont{and} \bibinfo{author}{\bibfnamefont{E.~V.}
  \bibnamefont{Sukhorukov}}, \bibinfo{journal}{Phys. Rev. B}
  \textbf{\bibinfo{volume}{78}}, \bibinfo{pages}{045322}
  (\bibinfo{year}{2008}).

\bibitem[{\citenamefont{le~Sueur et~al.}(2010)\citenamefont{le~Sueur,
  Altimiras, Gennser, Cavanna, Mailly, and
  Pierre}}]{le-Sueur-Altimiras-PRL-2010}
\bibinfo{author}{\bibfnamefont{H.}~\bibnamefont{le~Sueur}},
  \bibinfo{author}{\bibfnamefont{C.}~\bibnamefont{Altimiras}},
  \bibinfo{author}{\bibfnamefont{U.}~\bibnamefont{Gennser}},
  \bibinfo{author}{\bibfnamefont{A.}~\bibnamefont{Cavanna}},
  \bibinfo{author}{\bibfnamefont{D.}~\bibnamefont{Mailly}}, \bibnamefont{and}
  \bibinfo{author}{\bibfnamefont{F.}~\bibnamefont{Pierre}},
  \bibinfo{journal}{Phys. Rev. Lett.} \textbf{\bibinfo{volume}{105}},
  \bibinfo{pages}{056803} (\bibinfo{year}{2010}).

\bibitem[{\citenamefont{Altimiras
  et~al.}(2010{\natexlab{a}})\citenamefont{Altimiras, le~Sueur, Gennser,
  Cavanna, Mailly, and Pierre}}]{Altimiras-Nat-phys-2010}
\bibinfo{author}{\bibfnamefont{C.}~\bibnamefont{Altimiras}},
  \bibinfo{author}{\bibfnamefont{H.}~\bibnamefont{le~Sueur}},
  \bibinfo{author}{\bibfnamefont{U.}~\bibnamefont{Gennser}},
  \bibinfo{author}{\bibfnamefont{A.}~\bibnamefont{Cavanna}},
  \bibinfo{author}{\bibfnamefont{D.}~\bibnamefont{Mailly}}, \bibnamefont{and}
  \bibinfo{author}{\bibfnamefont{F.}~\bibnamefont{Pierre}},
  \bibinfo{journal}{Nature Physics} \textbf{\bibinfo{volume}{6}},
  \bibinfo{pages}{34} (\bibinfo{year}{2010}{\natexlab{a}}).

\bibitem[{\citenamefont{Altimiras
  et~al.}(2010{\natexlab{b}})\citenamefont{Altimiras, le~Sueur, Gennser,
  Cavanna, Mailly, and Pierre}}]{Altimiras-2010a}
\bibinfo{author}{\bibfnamefont{C.}~\bibnamefont{Altimiras}},
  \bibinfo{author}{\bibfnamefont{H.}~\bibnamefont{le~Sueur}},
  \bibinfo{author}{\bibfnamefont{U.}~\bibnamefont{Gennser}},
  \bibinfo{author}{\bibfnamefont{A.}~\bibnamefont{Cavanna}},
  \bibinfo{author}{\bibfnamefont{D.}~\bibnamefont{Mailly}}, \bibnamefont{and}
  \bibinfo{author}{\bibfnamefont{F.}~\bibnamefont{Pierre}},
  \bibinfo{journal}{Phys. Rev. Lett.} \textbf{\bibinfo{volume}{105}},
  \bibinfo{pages}{226804} (\bibinfo{year}{2010}{\natexlab{b}}).

\bibitem[{\citenamefont{Lunde et~al.}(2010)\citenamefont{Lunde, Nigg, and
  B\"uttiker}}]{Lunde-2010}
\bibinfo{author}{\bibfnamefont{A.~M.} \bibnamefont{Lunde}},
  \bibinfo{author}{\bibfnamefont{S.~E.} \bibnamefont{Nigg}}, \bibnamefont{and}
  \bibinfo{author}{\bibfnamefont{M.}~\bibnamefont{B\"uttiker}},
  \bibinfo{journal}{Phys. Rev. B} \textbf{\bibinfo{volume}{81}},
  \bibinfo{pages}{041311} (\bibinfo{year}{2010}).

\bibitem[{\citenamefont{Otsuka et~al.}(2010)\citenamefont{Otsuka, Abe, Iye, and
  Katsumoto}}]{Otsuka-PRB-2010}
\bibinfo{author}{\bibfnamefont{T.}~\bibnamefont{Otsuka}},
  \bibinfo{author}{\bibfnamefont{E.}~\bibnamefont{Abe}},
  \bibinfo{author}{\bibfnamefont{Y.}~\bibnamefont{Iye}}, \bibnamefont{and}
  \bibinfo{author}{\bibfnamefont{S.}~\bibnamefont{Katsumoto}},
  \bibinfo{journal}{Phys. Rev. B} \textbf{\bibinfo{volume}{81}},
  \bibinfo{pages}{245302} (\bibinfo{year}{2010}).

\bibitem[{\citenamefont{Degiovanni et~al.}(2010)\citenamefont{Degiovanni,
  Grenier, F\`eve, Altimiras, le~Sueur, and Pierre}}]{Degiovanni-PRB-2010}
\bibinfo{author}{\bibfnamefont{P.}~\bibnamefont{Degiovanni}},
  \bibinfo{author}{\bibfnamefont{C.}~\bibnamefont{Grenier}},
  \bibinfo{author}{\bibfnamefont{G.}~\bibnamefont{F\`eve}},
  \bibinfo{author}{\bibfnamefont{C.}~\bibnamefont{Altimiras}},
  \bibinfo{author}{\bibfnamefont{H.}~\bibnamefont{le~Sueur}}, \bibnamefont{and}
  \bibinfo{author}{\bibfnamefont{F.}~\bibnamefont{Pierre}},
  \bibinfo{journal}{Phys. Rev. B} \textbf{\bibinfo{volume}{81}},
  \bibinfo{pages}{121302} (\bibinfo{year}{2010}).

\bibitem[{\citenamefont{Kovrizhin and
  Chalker}(2011)}]{Kovrizhin-Chalker-PRB-2011}
\bibinfo{author}{\bibfnamefont{D.~L.} \bibnamefont{Kovrizhin}}
  \bibnamefont{and} \bibinfo{author}{\bibfnamefont{J.~T.}
  \bibnamefont{Chalker}}, \bibinfo{journal}{Phys. Rev. B}
  \textbf{\bibinfo{volume}{84}}, \bibinfo{pages}{085105}
  (\bibinfo{year}{2011}).

\bibitem[{\citenamefont{Levkivskyi and
  Sukhorukov}(2012)}]{Levkivskyi-Sukhorukov-PRB-2012}
\bibinfo{author}{\bibfnamefont{I.~P.} \bibnamefont{Levkivskyi}}
  \bibnamefont{and} \bibinfo{author}{\bibfnamefont{E.~V.}
  \bibnamefont{Sukhorukov}}, \bibinfo{journal}{Phys. Rev. B}
  \textbf{\bibinfo{volume}{85}}, \bibinfo{pages}{075309}
  (\bibinfo{year}{2012}).

\bibitem[{\citenamefont{Karzig et~al.}(2012)\citenamefont{Karzig, Levchenko,
  Glazman, and von Oppen}}]{Karzig-Levchenko-Glazman-Oppen-NJP-2012}
\bibinfo{author}{\bibfnamefont{T.}~\bibnamefont{Karzig}},
  \bibinfo{author}{\bibfnamefont{A.}~\bibnamefont{Levchenko}},
  \bibinfo{author}{\bibfnamefont{L.~I.} \bibnamefont{Glazman}},
  \bibnamefont{and} \bibinfo{author}{\bibfnamefont{F.}~\bibnamefont{von
  Oppen}}, \bibinfo{journal}{New J. of Phys.} \textbf{\bibinfo{volume}{14}},
  \bibinfo{pages}{105009} (\bibinfo{year}{2012}).

\bibitem[{\citenamefont{Chirolli et~al.}(2013)\citenamefont{Chirolli, Taddei,
  Fazio, and Giovannetti}}]{Chirolli-2013}
\bibinfo{author}{\bibfnamefont{L.}~\bibnamefont{Chirolli}},
  \bibinfo{author}{\bibfnamefont{F.}~\bibnamefont{Taddei}},
  \bibinfo{author}{\bibfnamefont{R.}~\bibnamefont{Fazio}}, \bibnamefont{and}
  \bibinfo{author}{\bibfnamefont{V.}~\bibnamefont{Giovannetti}},
  \bibinfo{journal}{Phys. Rev. Lett.} \textbf{\bibinfo{volume}{111}},
  \bibinfo{pages}{036801} (\bibinfo{year}{2013}).

\bibitem[{\citenamefont{Otsuka et~al.}(2014)\citenamefont{Otsuka, Sugihara,
  Yoneda, Nakajima, and Tarucha}}]{Otsuka-et-al-2014}
\bibinfo{author}{\bibfnamefont{T.}~\bibnamefont{Otsuka}},
  \bibinfo{author}{\bibfnamefont{Y.}~\bibnamefont{Sugihara}},
  \bibinfo{author}{\bibfnamefont{J.}~\bibnamefont{Yoneda}},
  \bibinfo{author}{\bibfnamefont{T.}~\bibnamefont{Nakajima}}, \bibnamefont{and}
  \bibinfo{author}{\bibfnamefont{S.}~\bibnamefont{Tarucha}},
  \bibinfo{journal}{J. of the Phys. Soc. of Japan}
  \textbf{\bibinfo{volume}{83}}, \bibinfo{pages}{014710}
  (\bibinfo{year}{2014}).

\bibitem[{\citenamefont{Inoue et~al.}(2014)\citenamefont{Inoue, Grivnin, Ofek,
  Neder, Heiblum, Umansky, and Mahalu}}]{Inoue-2014}
\bibinfo{author}{\bibfnamefont{H.}~\bibnamefont{Inoue}},
  \bibinfo{author}{\bibfnamefont{A.}~\bibnamefont{Grivnin}},
  \bibinfo{author}{\bibfnamefont{N.}~\bibnamefont{Ofek}},
  \bibinfo{author}{\bibfnamefont{I.}~\bibnamefont{Neder}},
  \bibinfo{author}{\bibfnamefont{M.}~\bibnamefont{Heiblum}},
  \bibinfo{author}{\bibfnamefont{V.}~\bibnamefont{Umansky}}, \bibnamefont{and}
  \bibinfo{author}{\bibfnamefont{D.}~\bibnamefont{Mahalu}},
  \bibinfo{journal}{Phys. Rev. Lett.} \textbf{\bibinfo{volume}{112}},
  \bibinfo{pages}{166801} (\bibinfo{year}{2014}).

\bibitem[{\citenamefont{Helzel et~al.}(2015)\citenamefont{Helzel, Litvin,
  Levkivskyi, Sukhorukov, Wegscheider, and Strunk}}]{Helzel-2015}
\bibinfo{author}{\bibfnamefont{A.}~\bibnamefont{Helzel}},
  \bibinfo{author}{\bibfnamefont{L.~V.} \bibnamefont{Litvin}},
  \bibinfo{author}{\bibfnamefont{I.~P.} \bibnamefont{Levkivskyi}},
  \bibinfo{author}{\bibfnamefont{E.~V.} \bibnamefont{Sukhorukov}},
  \bibinfo{author}{\bibfnamefont{W.}~\bibnamefont{Wegscheider}},
  \bibnamefont{and} \bibinfo{author}{\bibfnamefont{C.}~\bibnamefont{Strunk}},
  \bibinfo{journal}{Phys. Rev. B} \textbf{\bibinfo{volume}{91}},
  \bibinfo{pages}{245419} (\bibinfo{year}{2015}).

\bibitem[{\citenamefont{Tewari et~al.}(2016)\citenamefont{Tewari, Roulleau,
  Grenier, Portier, Cavanna, Gennser, Mailly, and Roche}}]{Tewari-2016}
\bibinfo{author}{\bibfnamefont{S.}~\bibnamefont{Tewari}},
  \bibinfo{author}{\bibfnamefont{P.}~\bibnamefont{Roulleau}},
  \bibinfo{author}{\bibfnamefont{C.}~\bibnamefont{Grenier}},
  \bibinfo{author}{\bibfnamefont{F.}~\bibnamefont{Portier}},
  \bibinfo{author}{\bibfnamefont{A.}~\bibnamefont{Cavanna}},
  \bibinfo{author}{\bibfnamefont{U.}~\bibnamefont{Gennser}},
  \bibinfo{author}{\bibfnamefont{D.}~\bibnamefont{Mailly}}, \bibnamefont{and}
  \bibinfo{author}{\bibfnamefont{P.}~\bibnamefont{Roche}},
  \bibinfo{journal}{Phys. Rev. B} \textbf{\bibinfo{volume}{93}},
  \bibinfo{pages}{035420} (\bibinfo{year}{2016}).

\bibitem[{\citenamefont{Slobodeniuk et~al.}(2016)\citenamefont{Slobodeniuk,
  Idrisov, and Sukhorukov}}]{Slobodeniuk-2016}
\bibinfo{author}{\bibfnamefont{A.~O.} \bibnamefont{Slobodeniuk}},
  \bibinfo{author}{\bibfnamefont{E.~G.} \bibnamefont{Idrisov}},
  \bibnamefont{and} \bibinfo{author}{\bibfnamefont{E.~V.}
  \bibnamefont{Sukhorukov}}, \bibinfo{journal}{Phys. Rev. B}
  \textbf{\bibinfo{volume}{93}}, \bibinfo{pages}{035421}
  (\bibinfo{year}{2016}).

\bibitem[{\citenamefont{Gurman et~al.}(2016)\citenamefont{Gurman, Sabo,
  Heiblum, Umansky, and Mahalu}}]{Gurman-2016}
\bibinfo{author}{\bibfnamefont{I.}~\bibnamefont{Gurman}},
  \bibinfo{author}{\bibfnamefont{R.}~\bibnamefont{Sabo}},
  \bibinfo{author}{\bibfnamefont{M.}~\bibnamefont{Heiblum}},
  \bibinfo{author}{\bibfnamefont{V.}~\bibnamefont{Umansky}}, \bibnamefont{and}
  \bibinfo{author}{\bibfnamefont{D.}~\bibnamefont{Mahalu}},
  \bibinfo{journal}{Phys. Rev. B} \textbf{\bibinfo{volume}{93}},
  \bibinfo{pages}{121412} (\bibinfo{year}{2016}).

\bibitem[{\citenamefont{Pascher et~al.}(2014)\citenamefont{Pascher, R\"ossler,
  Ihn, Ensslin, Reichl, and Wegscheider}}]{Pascher-2014}
\bibinfo{author}{\bibfnamefont{N.}~\bibnamefont{Pascher}},
  \bibinfo{author}{\bibfnamefont{C.}~\bibnamefont{R\"ossler}},
  \bibinfo{author}{\bibfnamefont{T.}~\bibnamefont{Ihn}},
  \bibinfo{author}{\bibfnamefont{K.}~\bibnamefont{Ensslin}},
  \bibinfo{author}{\bibfnamefont{C.}~\bibnamefont{Reichl}}, \bibnamefont{and}
  \bibinfo{author}{\bibfnamefont{W.}~\bibnamefont{Wegscheider}},
  \bibinfo{journal}{Phys. Rev. X} \textbf{\bibinfo{volume}{4}},
  \bibinfo{pages}{011014} (\bibinfo{year}{2014}).

\bibitem[{\citenamefont{Lunde and Nigg}(2016)}]{Lunde-2016}
\bibinfo{author}{\bibfnamefont{A.~M.} \bibnamefont{Lunde}} \bibnamefont{and}
  \bibinfo{author}{\bibfnamefont{S.~E.} \bibnamefont{Nigg}},
  \bibinfo{journal}{arxiv:1602.05039}  (\bibinfo{year}{2016}).

\bibitem[{\citenamefont{B{\"u}ttiker}(1986)}]{Buttiker-1986}
\bibinfo{author}{\bibfnamefont{M.}~\bibnamefont{B{\"u}ttiker}},
  \bibinfo{journal}{Phys. Rev. Lett.} \textbf{\bibinfo{volume}{57}},
  \bibinfo{pages}{1761} (\bibinfo{year}{1986}).

\bibitem[{\citenamefont{Blanter and B{\"u}ttiker}(2000)}]{Blanter2001}
\bibinfo{author}{\bibfnamefont{Y.}~\bibnamefont{Blanter}} \bibnamefont{and}
  \bibinfo{author}{\bibfnamefont{M.}~\bibnamefont{B{\"u}ttiker}},
  \bibinfo{journal}{Physics Reports} \textbf{\bibinfo{volume}{336}},
  \bibinfo{pages}{1 } (\bibinfo{year}{2000}).

\bibitem[{foo({\natexlab{a}})}]{footnote-1}
\bibinfo{note}{We neglect {\protect \em intra}-ES interaction, because the
  direct and exchange terms within each spin polarized ES tend to cancel each
  other~\cite {supp_mat,Lunde-2010,Lunde-2016}.}

\bibitem[{foo({\natexlab{b}})}]{footnote-2}
\bibinfo{note}{Since the slope (normal derivative) of the confinement potential
  increases as one approaches the edge, one typically expects $v_o > v_i$.}

\bibitem[{\citenamefont{Bocquillon et~al.}(2013)\citenamefont{Bocquillon,
  Freulon, Berroir, Degiovanni, Pla\c{c}ais, Cavanna, Jin, and
  F{\`e}ve}}]{Bocquillon-2013}
\bibinfo{author}{\bibfnamefont{E.}~\bibnamefont{Bocquillon}},
  \bibinfo{author}{\bibfnamefont{V.}~\bibnamefont{Freulon}},
  \bibinfo{author}{\bibfnamefont{J.-.~M.} \bibnamefont{Berroir}},
  \bibinfo{author}{\bibfnamefont{P.}~\bibnamefont{Degiovanni}},
  \bibinfo{author}{\bibfnamefont{B.}~\bibnamefont{Pla\c{c}ais}},
  \bibinfo{author}{\bibfnamefont{A.}~\bibnamefont{Cavanna}},
  \bibinfo{author}{\bibfnamefont{Y.}~\bibnamefont{Jin}}, \bibnamefont{and}
  \bibinfo{author}{\bibfnamefont{G.}~\bibnamefont{F{\`e}ve}},
  \bibinfo{journal}{Nat Commun} \textbf{\bibinfo{volume}{4}},
  \bibinfo{pages}{1839} (\bibinfo{year}{2013}).

\bibitem[{sup()}]{supp_mat}
\bibinfo{note}{See supplemental material at [URL provided by publisher]}.

\bibitem[{\citenamefont{Gardiner}(1983)}]{Gardiner}
\bibinfo{author}{\bibfnamefont{C.~W.} \bibnamefont{Gardiner}},
  \emph{\bibinfo{title}{Handbook of stochastic methods for physics, chemistry,
  and the natural sciences}} (\bibinfo{publisher}{Springer-Verlag, Berlin},
  \bibinfo{year}{1983}).

\bibitem[{foo({\natexlab{c}})}]{footnote-4}
\bibinfo{note}{When varying the path length difference instead of the flux, the
  given visibility is an approximation since the decay factor also depends on
  the path length difference. However, in the regime considered here, where
  $E_0-\mu_o\gg\Gamma_0, \eta L_i$, the oscillation period is much shorter than
  the characteristic decay length and this approximation is justified.}

\bibitem[{\citenamefont{Knuth}(1998)}]{Knuth-1998}
\bibinfo{author}{\bibfnamefont{D.~E.} \bibnamefont{Knuth}},
  \emph{\bibinfo{title}{The Art of Computer Programming}}
  (\bibinfo{publisher}{Addison-Wesley}, \bibinfo{year}{1998}),
  \bibinfo{edition}{3rd} ed.

\bibitem[{\citenamefont{Beenakker}(1991)}]{Beenakker-1991}
\bibinfo{author}{\bibfnamefont{C.~W.~J.} \bibnamefont{Beenakker}},
  \bibinfo{journal}{Phys. Rev. B} \textbf{\bibinfo{volume}{44}},
  \bibinfo{pages}{1646} (\bibinfo{year}{1991}).

\end{thebibliography}
\bibliographystyle{apsrev}

\cleardoublepage

\begin{widetext}
\begin{center}
{\bf Supplementary Material for\\``Stochastic theory of
  interaction-induced decoherence of weakly disordered edge states''}\vspace{0.5cm}\\
{Simon E. Nigg and Anders Mathias Lunde}\vspace{0.5cm}
\begin{minipage}[c]{0.76\textwidth}\vspace{0.2cm}
{\small \hspace{0.24cm} This supplementary material contains further information to complement
the main text. In particular, we provide: 1) An explicit model for the
kinetic Boltzmann equation including momentum conservation breaking
disorder. 2) A physically motivated Monte Carlo
simulation of the energy exchange dynamics based on the kinetic
Boltzmann equation. 3) A derivation of the identity (15) of the main
text. 4) Details on the parameter estimation from the experimental data of
Tewari et al.~[\onlinecite{Tewari-2016}].}
\end{minipage}
\end{center}

\section{1. ~Effective Coulomb interaction kernel for disordered edge states}
While edge states of the integer quantum
Hall effect are often described as translationally invariant one
dimensional channels on length scales of hundreds of microns, recent scanning tunneling experiments in high
mobility samples have detected edge
roughness with a characteristic length scale of a few hundred
nanometers~[\onlinecite{Pascher-2014}].

Here we derive an effective Coulomb interaction kernel for inter-ES
scattering, in the presence
of translation invariance breaking disorder, which has the form used
in the collision integral Eq.~(7) of the main paper. This extends the earlier derivations given in~[\onlinecite{Lunde-2010, Lunde-2016}].

Our starting point is the Boltzmann kinetic equation for the
distribution functions of the inner and outer ESs in momentum space
\begin{align}\label{eqSM:6}
v_{\alpha}\partial_xf_{\alpha,k}(x)&=\sum_{k_2k_{1'}k_{2'}}W_{12,1'2'}\Big\{f_{1'}(x)[1-f_{1}(x)]f_{2'}(x)[1-f_{2}(x)]
-f_{1}(x)[1-f_{1'}(x)]f_{2}(x)[1-f_{2'}(x)]\Big\},
\end{align}
where we use the shorthand notation 
$1=\alpha,k_1$ and $2=\bar\alpha,k_2$ and $\alpha\in\{i,o\}$ and
$\bar\alpha=o\delta_{\alpha i} + i\delta_{\alpha o}$. The interaction kernel is obtained via
Fermi's golden rule and reads
\begin{align}
  W_{12,1'2'}=\frac{2\pi}{\hbar}|\braket{k_{1'}\alpha,k_{2'}\bar\alpha|V|k_1\alpha,k_2\bar\alpha}|^2\delta(E_{k_1\alpha}+E_{k_2\bar\alpha}-E_{k_{1'}\alpha}-E_{k_{2'}\bar\alpha}),
\end{align}
The delta
function in energy enforces energy conservation during the
collision. Here $V$ is the Coulomb interaction operator, the relevant matrix
elements for inter-ES scattering of which are given by
\begin{align}
V_{12,1'2'}&\equiv\braket{k_{1'}\alpha,k_{2'}\bar\alpha|V|k_1\alpha,k_2\bar\alpha}\nonumber\\
&=\int dx_1dx_2\int dy_1dy_2\psi_{1'}^*(x_1,y_1)\psi_{2'}^*(x_2,y_2)V_C(x_1-x_2,y_1-y_2)\psi_1(x_1,y_1)\psi_2(x_2,y_2).\label{eqSM:5}
\end{align}
Here
\begin{align}
\psi_{k,\alpha}(x,y)=\frac{1}{\sqrt{L}}e^{ik_{\alpha}x}\frac{1}{\pi^{1/4}\sqrt{\ell_B}}\exp\left[ -\frac{(y-y_{\alpha}(x))^2}{2\ell_B^2} \right],
\end{align}
is the wavefunction of an electron in ES $\alpha$, which is localized
in the transverse direction to within the magnetic length
$\ell_B=\sqrt{\hbar/(|e|B)}$, on the guiding center coordinate
$y_{\alpha}(x)$. Disorder will be included in the dependence of the guiding center
coordinate on the longitudinal coordinate $x$, as explained further below. Note that we suppress the spin index,
keeping in mind that the two edge states have opposite spins. Because
of this, the exchange term is absent for inter-ES interaction. For
{\em intra}-ES interaction, both direct and exchange terms are present
(since the ESs are spin polarized), and typically compensate each
other. In a model with contact interaction, such as used below, this
cancellation is complete, resulting in a vanishing matrix element. In general, the cancellation is not
exact, but for sufficiently short range interaction, the inter-ES interaction dominates over the intra-ES
interaction, justifying neglecting the latter~\cite{Lunde-2010,Lunde-2016}. $L$ denotes the length
of the system in the propagation direction. The Coulomb potential energy
in the plane of the 2DEG at $z=0$ is
\begin{align}
V_C(x_1-x_2,y_1-y_2)=\frac{e^2}{4\pi\epsilon_0}\frac{e^{-\sqrt{(x_1-x_2)^2+(y_1-y_2)^2}/\ell_s}}{\sqrt{(x_1-x_2)^2+(y_1-y_2)^2}}.
\end{align}
Here we include screening by the metallic gates with characteristic
screening length $\ell_s$. Substituting into Eq.~(\ref{eqSM:5}) yields
\begin{align}
V_{121'2'}&=\frac{1}{L^2}\frac{e^2}{4\pi\epsilon_0}\int
  dx_1dx_2e^{i(k_1-k_{1'})x_1+i(k_2-k_{2'})x_2}\nonumber\\
&\times\frac{1}{\pi\ell_B^2}\int dy_1dy_2\frac{e^{-\sqrt{(x_1-x_2)^2+(y_1-y_2)^2}/\ell_s}}{\sqrt{(x_1-x_2)^2+(y_1-y_2)^2}}\exp\left[ -\frac{[y_1-y_{\alpha}(x_1)]^2}{\ell_B^2} \right]\exp\left[-\frac{[y_2-y_{\bar\alpha}(x_2)]^2}{\ell_B^2}  \right].
\end{align}
To make further progress we now assume that $\ell_B\ll\ell_s$, in which case we can
approximate the Gaussians by delta functions and perform the integrals
over $y_1$ and $y_2$ to obtain
\begin{align}
V_{121'2'}&\simeq\frac{1}{L^2}\frac{e^2}{4\pi\epsilon_0}\int
  dx_1dx_2e^{i(k_1-k_{1'})x_1+i(k_2-k_{2'})x_2}\frac{e^{-\sqrt{(x_1-x_2)^2+[y_{\alpha}(x_1)-y_{\bar\alpha}(x_2)]^2}/\ell_s}}{\sqrt{(x_1-x_2)^2+[y_{\alpha}(x_1)-y_{\bar\alpha}(x_2)]^2}}.
\end{align}
Next, we introduce the relative and center of mass coordinates
$r=(x_1-x_2)/2$ and $R=(x_1+x_2)/2$, and write
$y_{\alpha}(x_1)-y_{\bar\alpha}(x_2)=\Delta y_0+\delta y(r,R)$, where
the disorder induced deviation is small in the sense that $|\delta y(r,R)|\ll \Delta
y_0$. Then
\begin{align}
V_{121'2'}&=\frac{1}{L^2}\frac{e^2}{4\pi\epsilon_0}\int
  drdRe^{i\Delta qr+i\Delta k
            R}\frac{e^{-\sqrt{(2r)^2+[\Delta y_0+\delta
            y(r,R)]^2}/\ell_s}}{\sqrt{(2r)^2+[\Delta y_0+\delta y(r,R)]^2}},
\end{align}
with $\Delta q=k_1-k_2-k_{1'}+k_{2'}$ and $\Delta k=k_1+k_2
-k_{1'}-k_{2'}$. If we assume that the momentum exchanges are small,
in the sense that $\Delta q \ell_s\ll 1$, then we can approximate the
integral over $r$ by $\ell_s$ times the integrand at $r=0$, which
yields
\begin{align}
V_{121'2'}&\simeq\frac{\ell_s}{L^2}\frac{e^2}{4\pi\epsilon_0}\int
  dRe^{i\Delta k
            R}\frac{e^{-[\Delta y_0+\delta
            y(R)]/\ell_s}}{\Delta y_0+\delta y(R)}\nonumber\\
&\simeq\frac{1}{L^2}\frac{e^2}{4\pi\epsilon_0}\frac{\ell_se^{-\Delta y_0/\ell_s}}{\Delta y_0}\int
  dRe^{i\Delta k
            R}(1-\delta
            y(R)/\ell_s),
\end{align}
where $\delta y(R)=\delta y(0,R)$ and, in the last step, we have
assumed that $|\delta y(R)|\ll\ell_s$. Thus the matrix elements splits
into the sum of two contributions. The first one, which is
proportional to $\frac{1}{L}\int dRe^{i\Delta kR}=\delta_{0\Delta k}$,
represents the momentum conserving part of the scattering. Because, for
linear dispersion with different ES velocities, inelastic collisions
conserving both momentum and energy are forbidden, this term does not
contribute to the kernel. The second contribution to the matrix
element is proportional to $\int dRe^{i\Delta kR}\delta y(R)$ and its
contribution to the kernel will
in general not vanish in the presence of disorder (i.e. for $\delta y
(R)\not=\rm const$). The function $\delta y(R)$
depends on the particular realization of disorder. Since we are not
interested in a specific disorder realization, we model its effect
by assuming Gaussian correlated fluctuations, i.e.
\begin{align}
\braket{\delta y(R)}_{\rm disorder}=0,\quad\braket{\delta y(R)\delta
  y(R')}_{\rm disorder}=\frac{A}{\sqrt{2\pi}\ell_p}\exp\left[ -\frac{(R-R')^2}{2\ell_p^2} \right].
\end{align}
Here $\sqrt{A/\ell_p}$ determines the maximum magnitude of the
transverse fluctuations while the
momentum conservation breaking correlation length
$\ell_p$, characterizes the edge roughness in the propagation
direction ($\ell_p\rightarrow\infty$ for a translation invariant
system). With this model of disorder, we can now compute the relevant
disorder averaged squared matrix element for non-momentum conserving scattering
\begin{align}
\braket{|V_{121'2'}^{(\Delta k\not=0)}|^2}_{\rm disorder}=\left(
  \frac{1}{L^2}\frac{e^2}{4\pi\epsilon_0}\frac{e^{-\Delta
  y_0/\ell_s}}{\Delta y_0}\right)^2\frac{A}{\sqrt{2\pi}\ell_p}\int
  dRdR'e^{i\Delta k(
            R-R')}\exp\left[ -\frac{(R-R')^2}{2\ell_p^2}
  \right]\simeq \frac{A}{L^3}\left( \frac{e^2}{4\pi\epsilon_0}\frac{e^{-\Delta
  y_0/\ell_s}}{\Delta y_0}\right)^2e^{ -(\Delta k\ell_p)^2/2}.
\end{align}
Using the dispersion relation
$k_{\alpha}=\frac{E_{\alpha}}{\hbar v_{\alpha}}$ and energy
conservation $E_1+E_2-E_{1'}-E_{2'}=0$, we can write $\Delta
k\ell_p/\sqrt{2}=\omega/\Delta E$, with
\begin{align}
\omega &= E_1-E_{1'}=E_{2'}-E_{2},\\
\Delta E&=\frac{\sqrt{2}}{\ell_p}\frac{\hbar v_iv_o}{|v_i-v_o|}.
\end{align}
Substituting into Eq.~(\ref{eqSM:6}) and changing from discrete momentum summation to continuum integration over
energy ($\sum_{k_{\alpha}}\rightarrow \frac{L}{\hbar v_{\alpha}}\int
dE$), we finally obtain after some algebra
\begin{align}\label{eqSM:7}
\partial_xf_{\alpha}(x) = \gamma\int_{-\infty}^{\infty}d\omega
  e^{-\left( \omega/\Delta E\right)^2}\Big\{ f_{\alpha}(E+\omega,x)[1-f_{\alpha}(E,x)]D_{\bar\alpha}(\omega,x)-f_{\alpha}(E,x)[1-f_{\alpha}(E+\omega,x)]D_{\bar\alpha}(-\omega,x) \Big\},
\end{align}
with
\begin{align}\label{eqSM:8}
\gamma = \frac{2\pi A}{(\hbar v_{i}\hbar v_{o})^2}V_0^2,\quad
  V_0=\frac{e^2}{4\pi\epsilon_0}\frac{e^{-\Delta y_0/\ell_s}}{\Delta y_0},
\end{align}
and
$D_{\alpha}(\omega,x)=\int_{-\infty}^{\infty}dE'f_{\alpha}(E'-\omega,x)[1-f_{\alpha}(E',x)]$.
In conclusion, we have derived an effective model for Coulomb
interaction between two weakly disordered ESs. A few comments to
conclude: Firstly, the divergence of $\Delta E$ for $v_i=v_o$ is an artifact of
using a linear dispersion relation. A linear dispersion relation is not essential
for the momentum conservation breaking physics but is convenient for
computations. Furthermore, in general one can expect that $v_o>v_i$,
since the outer ES is closer to the edge of the sample. Secondly, we
note that $V_0\rightarrow 0$ for $\ell_s\ll\Delta y_0$. This is
intuitively reasonable, since the edge states are separated by a
finite distance $\sim\Delta y_0$ and if the screening is too strong,
electrons on the inner and outer ESs do not interact. Finally, we note
that Eq.~(\ref{eqSM:7}) can be
obtained directly~\cite{Lunde-2010, Lunde-2016}, by starting with an
effective one-dimensional local inter-ES
interaction potential of the form $V_{\rm eff}(x,x')=V_0g(x)\delta(x-x')$, with
$\braket{(g(x)-g_0)(g(x')-g_0)}=A/(\sqrt{2\pi}\ell_p)\exp[-(x-x')^2/(2\ell_p^2)]$
and $g_0=\braket{g(x)}$.

\section{2.~Monte Carlo simulation}
In this section we describe a Monte Carlo simulation of the
Fokker-Planck dynamics described by the kinetic Eq.~(6) and (7) of the
main text.
This simulation is based on discretizing the stochastic energy
exchange process in space. To form a qualitative picture, imagine
following an injected electron as it propagates along the outer ES. At
random times $t_i=v_o x_i$, it will scatter off an electron in the
inner ES, changing its energy and thereby the phase accumulation
rate. The phase at a given distance is then a random
number given by the sum of the phases accumulated in every
interval up to that distance. Importantly, both the scattering rate
and the scattering energy probability distribution depend on position
and need to be updated in each interval. Next we describe in detail
how this is achieved.

We want to compute numerically the expectation value of a function
$\xi(\phi_E)$ of
the phase $\phi_E$ accumulated by an electron, with initial energy $E$, propagating in the outer ES along
one arm of the MZI. Let $L$
denote the total arm length. From the kinetic Eq.~(6) and (7) of
the main text, the number of scattering events per length for an
electron at position $x$ is
\begin{align}
S(x) = \gamma\int_{-\infty}^{\infty}d\omega e^{-\left(
  \frac{\omega}{\Delta E} \right)^2}\left[ 1-b(E,x)
  \right]D_i(\omega,x)\simeq \gamma\int_{-\infty}^{\infty}d\omega e^{-\left(
  \frac{\omega}{\Delta E}
  \right)^2}D_i(\omega,x)\simeq\gamma\sqrt{\pi}\Delta E D_i(0,x),
\end{align}
where the middle expression holds in the Fokker-Planck limit where
$b(E,x)\ll 1$ and the last expression holds for weak momentum
conservation breaking, where $\Delta E$ is the smallest energy scale. 

It can further be easily shown that
the scattering probability density from energy $E$ to $E'$ at position $x$ is
\begin{align}\label{eqSM:1}
P_{E\rightarrow E'}=\frac{e^{-\left( \frac{E-E'}{\Delta E}
  \right)^2}D_i(E-E',x)}{\int_{-\infty}^{\infty}dE'e^{-\left(
  \frac{E-E'}{\Delta E} \right)^2}D_i(E-E',x)}.
\end{align}
Finally, within the effective temperature approximation (see Eq.~(20) of
the main text)
\begin{align}
D_i(E-E',x)\simeq \frac{E-E'}{1-\exp\left[ -\frac{E-E'}{\kb T_i(x)} \right]}.
\end{align}

The Monte-Carlo algorithm we implement is now as follows:
We discretize the length of the interferometer arm into segments of size $\Delta x=L/N$ with some
suitably large integer $N$ such that $\Delta x S(x)< 1$. Then,
$\Delta x S(x)$ gives the scattering probability in the interval
$[x,x+\Delta x]$. Using
the Metropolis-Hastings algorithm, in
every interval $n=0\dots N-1$, we draw a uniformly distributed random number $r_n\in[0,1)$ and
if $\Delta x S(n\Delta x) > r_n$, we scatter the electron's energy
by drawing a
random number $E'$ from the distribution $P_{E_n\rightarrow E'}$,
i.e. $E((n+1)\Delta x)= E'$, otherwise we leave
the energy unchanged, i.e. $E((n+1)\Delta x)=E(n\Delta x)\equiv E_n$. The
accumulated phase for one such ``trajectory'' from $x=0$ to $x=N\Delta
x=L$ is then approximated by
\begin{align}
\phi_E=\frac{\Delta x}{\hbar v_o}\sum_{n=0}^{N-1}E(n\Delta x).
\end{align}
The injection energy is given by the initial condition $E(0)=E$. We repeat this loop $M$ times and estimate the sample average and variance of $\xi(\phi)$
according to Knuth's online algorithm (See. e.g.~[\onlinecite{Knuth-1998}]).
In Fig.~(2) of the main text
, we show the resulting visibility for an
interferometer with equal arm lengths $x$, $\mathcal{V}(x)=|\mathcal{F}_E(x)|^2=|\frac{1}{M}\sum_m\exp(i\phi_{E,m}(x))|^2$, for different values of the injection
energy and $M=40000$ trajectories per energy. The Monte Carlo simulation results
perfectly confirm
our analytic predictions to within the statistical uncertainty $\sim
1/\sqrt{M}=0.5\%$.

The code of our implementation is written in python 3 using the
numerical libraries numpy and scipy and is made available for
inspection upon request. Please send inquiries to \href{mailto:simon.nigg@unibas.ch}{simon.nigg@unibas.ch}.

\section{3.~Double integral with stochastic term}
Here we prove Eq.~(15) of the main
text. Consider the stochastic integral
\begin{align}
K_y=\int_0^yg(z)\dd W_z.
\end{align}
Using It\^o's calculus
we have
\begin{align}
d(K_yy)=\dd K_y y + K_ydy + \dd K_y dy.
\end{align}
Because $\dd K_y=g(y)\dd W_y$ and since $\dd W_ydy=0$, we simply have,
as in normal calculus
\begin{align}
\int_0^x\int_0^yg(z)\dd W_zdy=\int_0^xK_ydy=\int_0^x\left[ d(K_yy)-\dd K_yy \right]=K_x x - \int_0^xyg(y)\dd
  W_y=\int_0^x(x-y)g(y)\dd W_y.
\end{align}

\section{4.~Estimation of parameters from experimental data}
Here we explain in more details how we estimated the parameters $\eta$
and $N_b/\rho_o^{}$ from the experimental data presented
in~[\onlinecite{Tewari-2016}]. The transmission probability of the injection
QD is given in Fig. 2(b) of~[\onlinecite{Tewari-2016}] and fits with that
of a thermally broadened QD level given by~[\onlinecite{Beenakker-1991}]
\begin{align}
P_{\rm transmission}=H\cosh^{-2}\left( \frac{E_0-\mu_o^{}}{2\kb T} \right),
\end{align}
with electronic temperature $T\approx 31\,{\rm mK}$.
$E_0-\mu_o^{}$ is the energy difference between the QD
energy level (average injection energy) and the Fermi energy of the
outer ES. The maximal measured transmission probability is $H\approx 0.15$. The parameter $N_b/\rho_o$ is then simply given
by the integral of the transmission curve, that is
\begin{align}
\frac{N_b}{\rho_o^{}}=\int dE_0P_{\rm transmission}=4H\kb T\approx1.6\,{\rm \mu eV}.
\end{align}
Having determined $N_b/\rho_o^{}$, the only remaining free parameter
is $\eta$. An estimate for the latter is obtained by fitting the
visibility as follows. Since the interferometer used in~[\onlinecite{Tewari-2016}]
is approximately symmetric, we assume equal arm lengths $\Delta L=0$. Furthermore, in the experiment the QPCs are
tuned to be semi-transparent,
i.e. $r_1=r_2=t_1=t_2=1/\sqrt{2}$. According to Eq.(19) of the main
text, the
visibility is then simply given by the absolute value squared of the coherence suppression factor i.e.
\begin{align}\label{eqSM:2}
\mathcal{V}(x)=\left|\mathcal{F}_E(\phi)\right|^2=e^{-\delta\phi^2(x)}.
\end{align}
The variance of the phase is given by Eq.~(16) of the main text and in the
effective temperature approximation is
\begin{align}\label{eqSM:3}
\delta\phi^2(x)=\frac{2\eta\kb}{(\hbar v_o)^2}\int_0^x(x-y)^2T_i(y)dy.
\end{align}
Here, the effective temperature $T_i(x)$ is determined by the system of
differential equations~(22) of the main text and thereby depends on $\eta$ and $N_b/\rho_o^{}$.

In the energy range $30\,{\rm \mu eV}<E_0-\mu_o^{}<120\,{\rm \mu eV}$
the measured visibility is independent of the injection energy and its
mean value is approximately~[\onlinecite{Tewari-2016}]
\begin{align}\label{eqSM:4}
\mathcal{V}(L)\approx 0.125,
\end{align}
where $L\approx 7.2\,{\rm \mu m}$, is the arm length of the
interferometer (See caption of Fig. 1 in~[\onlinecite{Tewari-2016})].
\begin{figure}[ht]
\includegraphics[width=0.5\textwidth]{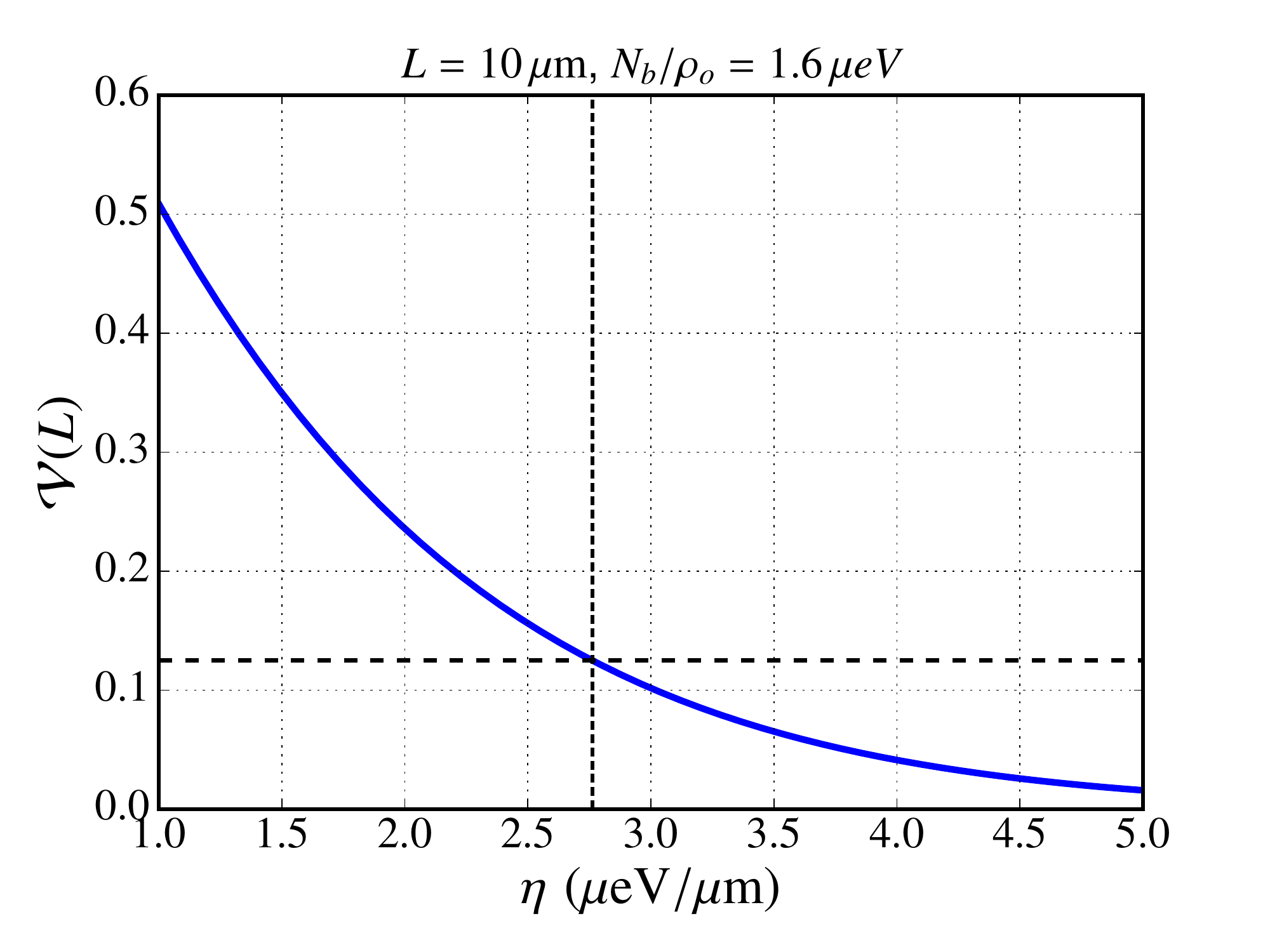}\caption{Graphical
  determination of the energy drift velocity. The (blue) solid curve
  is computed by solving numerically the differential equations~(21)
  of the main text for the effective temperature $T_i(x)$. The horizontal (black)
  dashed line gives the value of the visibility measured
  in~[\onlinecite{Tewari-2016}] and the vertical (black) dashed line
  indicates the fitting value of $\eta$.\label{fig:eta}}
\end{figure}
We compute numerically $\mathcal{V}(L)$ according to Eqs.~(\ref{eqSM:2})
and~(\ref{eqSM:3}) and plot the solution as a function of $\eta$ in
Fig.~\ref{fig:eta}. Comparing with Eq.~(\ref{eqSM:4}) we find that the value of the energy
drift velocity consistent with the experiment is
\begin{align}
\eta\approx 2.8\,\frac{\rm \mu eV}{\rm \mu m}.
\end{align}

With these parameters thus determined, we find for the crossover
distance defined in the main text $x_s=(\pi
\kb T)^2\rho_o^{}/(3\eta N_b)\approx 5.2\,{\rm \mu m}$.

\end{widetext}

\end{document}